\documentclass[%
 reprint,
superscriptaddress,
 amsmath,amssymb,
 aps,
floatfix,
showkeys
]{revtex4-2}

\usepackage{graphicx}% Include figure files
\usepackage{dcolumn}% Align table columns on decimal point
\usepackage{bm}% bold math
\usepackage[utf8]{inputenc}
\usepackage{tabularx}
\usepackage[dvipsnames]{xcolor}
\usepackage{units}
\usepackage[euler]{textgreek}
\usepackage{upgreek}
\usepackage{derivative}
\usepackage{amsmath}
\usepackage{physics}
\usepackage{amssymb}
\usepackage[colorlinks=true,linkcolor=blue!50!black,urlcolor=blue!50!black,citecolor=blue!50!black]{hyperref}
\usepackage{afterpage}
\usepackage{subfigure}
\usepackage{natbib}
\usepackage{xurl}

\DeclareGraphicsExtensions{.pdf,.png,.jpg}

\begin{document}

\preprint{APS}

%!TEX root = main.tex
\newcommand{\mpi}{\affiliation{Max-Planck-Institut f\"ur Physik, 85748 Garching bei M\"unchen, Germany}}
\newcommand{\coimbra}{\affiliation{Also at: LIBPhys, Departamento de Fisica, Universidade de Coimbra, P3004 516 Coimbra, Portugal}}
\newcommand{\hephy}{\affiliation{Institut f\"ur Hochenergiephysik der \"Osterreichischen Akademie der Wissenschaften, 1010 Wien, Austria}}
\newcommand{\ati}{\affiliation{Atominstitut, Technische Universit\"at Wien, 1020 Wien, Austria}}
\newcommand{\tum}{\affiliation{Physik-Department, TUM School of Natural Sciences, Technische Universit\"at M\"unchen, 85747 Garching, Germany}}
\newcommand{\tuebingen}{\affiliation{Eberhard-Karls-Universit\"at T\"ubingen, 72076 T\"ubingen, Germany}} 
\newcommand{\bratislava}{\affiliation{Faculty of Mathematics, Physics and Informatics, Comenius University, 84248 Bratislava, Slovakia}}

\newcommand{\oxford}{\affiliation{Department of Physics, University of Oxford, Oxford OX1 3RH, United Kingdom}}
\newcommand{\wmi}{\affiliation{Also at: Walther-Mei\ss ner-Institut f\"ur Tieftemperaturforschung, 85748 Garching, Germany}}
\newcommand{\mib}{\affiliation{Also at: Dipartimento di Fisica, Università di Milano Bicocca, Milano, 20126, Italy}}
\newcommand{\lngs}{\affiliation{INFN, Laboratori Nazionali del Gran Sasso, 67010 Assergi, Italy}}
\newcommand{\cassino}{\affiliation{Also at: Dipartimento di Ingegneria Civile e Meccanica, Università degli Studi di Cassino e del Lazio Meridionale, 03043 Cassino, Italy}}
\newcommand{\uhd}{\affiliation{Kirchhoff-Institute for Physics, Heidelberg University, 69120 Heidelberg, Germany}}
\newcommand{\usp}{\affiliation{Also at: Instituto de Física da Universidade de São Paulo, São Paulo 05508-090, Brazil}}
\newcommand{\kit}{\affiliation{Institute for Astroparticle Physics, Karlsruhe Institute of Technology, 76128 Karlsruhe, Germany}}
\newcommand{\melb}{\affiliation{Now at: School of Physics, The University of Melbourne, Melbourne, VIC 3010, Australia}}
\newcommand{\arc}{\affiliation{Now at: ARC Centre of Excellence for Dark Matter Particle Physics, Australia}}

\mpi
\hephy
\ati
\bratislava
\lngs
\tum
\uhd
\kit
\tuebingen
\oxford
\coimbra
\mib
\wmi
\usp
\cassino

\author{G.~Angloher}
  \mpi

\author{S.~Banik}
  \hephy
  \ati

\author{A.~Bento}
  \mpi
  \coimbra 

\author{A.~Bertolini}
  \uhd

\author{R.~Breier}
  \bratislava

\author{C.~Bucci}
  \lngs 

\author{J.~Burkhart}
  \hephy

\author{L.~Canonica}
  \mpi
  \mib

\author{E.R.~Cipelli}
  \mpi

\author{S.~Di~Lorenzo}
  \mpi
  \lngs

\author{J.~Dohm}
  \tuebingen 

\author{F.~Dominsky} \email[Corresponding author: ]{dominsky@mpp.mpg.de}
  \mpi

\author{L.~Einfalt}
  \hephy
  \ati
  \melb
  \arc
  
\author{A.~Erb}
  \tum
  \wmi
  
\author{E.~Fascione}   
    \uhd

\author{F.~v.~Feilitzsch}
  \tum 
   
 \author{S.~Fichtinger}
  \hephy

\author{D.~Fuchs}
   \hephy
   \ati
   
 \author{V.M.~Ghete}
  \hephy 

\author{P.~Gorla}
  \lngs 

\author{P.V.~Guillaumon}\email[Corresponding author: ]{pedro.guillaumon@mpp.mpg.de}
  \mpi
  \usp

\author{D.~Hauff}
  \mpi 

\author{M.~Ješkovsk\'y}
  \bratislava

\author{J.~Jochum}
  \tuebingen 

\author{M.~Kaznacheeva}
  \tum
  
\author{H.~Kluck}
  \hephy

\author{H.~Kraus}
  \oxford

\author{B.~v.~Krosigk} \email[Corresponding author:\\ ]{bkrosigk@kip.uni-heidelberg.de}
    \uhd
    \kit

\author{A.~Langenk\"amper}
  \mpi

\author{M.~Mancuso}
  \mpi

\author{B.~Mauri}
  \mpi

\author{V.~Mokina}\email[Corresponding author:\\]{valentyna.mokina@oeaw.ac.at}
  \hephy

\author{C.~Moore}
  \mpi

\author{P.~Murali}
    \uhd

\author{M.~Olmi}
  \lngs
  
\author{T.~Ortmann}
  \tum

\author{C.~Pagliarone}
  \lngs 
  \cassino

\author{L.~Pattavina}
  \lngs
  \mib

\author{F.~Petricca}
  \mpi 

\author{W.~Potzel}
  \tum 

\author{P.~Povinec}
  \bratislava

\author{F.~Pr\"obst}
  \mpi

\author{F.~Pucci}
  \lngs
  
\author{F.~Reindl}
  \hephy
  \ati

\author{J.~Rothe}
  \tum
  
\author{K.~Sch\"affner}
  \mpi

\author{J.~Schieck}
  \hephy
  \ati 

\author{S.~Sch\"onert}
  \tum 
  
\author{C.~Schwertner}
  \hephy
  \ati

\author{M.~Stahlberg}
  \mpi

\author{L.~Stodolsky}
  \mpi 

\author{C.~Strandhagen}
  \tuebingen

\author{R.~Strauss}
  \tum

\author{I.~Usherov}
  \tuebingen 
  
\author{D.~Valdenaire}
 \hephy
 \ati

\author{M.~Zanirato}
  \mpi
  
\author{V.~Zema}
  \mpi
  \hephy
  
\collaboration{CRESST Collaboration}
\noaffiliation

\title{The CRESST experiment: towards the next-generation of sub-GeV direct dark matter detection}

%\input{authors.tex}
%\thanks{Update with a working title}
%\date{\today}

\begin{abstract}
Direct detection experiments have established the most stringent constraints on potential interactions between particle candidates for relic, thermal dark matter and Standard Model particles. To surpass current exclusion limits a new generation of experiments is being developed.
The upcoming upgrade of the CRESST experiment will incorporate $\mathcal{O}$(100) detectors with different masses ranging from $\sim$2\,g to $\sim$24\,g, aiming to achieve unprecedented sensitivity to sub-GeV dark matter particles with a focus on spin-independent dark matter-nucleus scattering.
This paper presents a comprehensive analysis of the planned upgrade, detailed experimental strategies, anticipated challenges, and projected sensitivities. Approaches to address and mitigate low-energy excess backgrounds — a key limitation in previous and current sub-GeV dark matter searches — are also discussed. In addition, a long-term roadmap for the next decade is outlined, including other potential scientific applications.

\end{abstract}


\maketitle

\section{Introduction}\label{intro}

Since the 1930s, significant efforts have been made to understand the nature of dark matter (DM) with experimental searches for DM particles starting to gain momentum in the 1980s. 
DM particle candidates span a mass range of about 50 orders of magnitude, from about $10^{-22}$\,eV up to the Planck scale. 
For decades, weakly interacting massive particles (WIMPs), generated thermally during the hot, early universe, have been one of the most studied DM candidates \sloppy\cite{10.21468/SciPostPhysLectNotes.71}.
Of these, the most prominent role over the years has been played by a WIMP with a mass above a few GeV/$c^2$ and below a few hundred TeV/$c^2$, often referred to as standard WIMP \cite{PhysRevLett.39.165, Smirnov_2019}. 
However, theoretically well-motivated WIMP models with sub-GeV masses also exist, commonly called light DM. 
The latter have been gaining increasing attention in recent years due to the lack of a standard WIMP observation to date \cite{Lin:2019uvt, cirelli2024darkmatter}.
Over the past 40 years, a variety of approaches, techniques, and experiments have been employed to tackle this complex puzzle. 
Among them, the Cryogenic Rare Event Search with Superconducting Thermometers (CRESST) experiment has been a key player for over 20 years, advancing the direct search for WIMPs with masses of a few GeV/$c^2$ and later extending below 1\,GeV/$c^2$, using cryogenic calorimetry.

The CRESST experiment began acquiring data in the Laboratori Nazionali del Gran Sasso (LNGS, Italy) in 2000, initially using sapphire (Al$_2$O$_3$) crystals \cite{Angloher:2002in}.
Currently in its third phase, the experiment operates cryogenic calorimeters made from crystalline materials such as Al$_2$O$_3$, CaWO$_4$ \cite{CRESST:2020wtj}, Si \cite{PhysRevD.107.122003}, and LiAlO$_2$ \cite{CRESST:2022dtl} equipped with highly sensitive transition-edge sensors (TESs)  operated at approximately 15\,mK. 
The main goal of the experiment is to discover DM, an effort that is still ongoing.
In the meantime, CRESST has been able to probe DM candidates with spin-independent DM-nucleon scattering cross sections down to $\mathcal{O}(10^{-42})\,\text{ cm}^2$ at 10\,GeV/$c^2$~\cite{CRESST:2019jnq}.
Thanks to the low thresholds achieved by this technology, as low as 6.7\,eV for gram-scale detectors \cite{CRESST:2024cpr} and 30.1\,eV for a 23.6\,g  crystal~\cite{CRESST:2019jnq}, CRESST has additionally successfully probed sub-GeV DM with spin-independent DM-nucleon scattering cross sections down to $\mathcal{O}(10^{-38})\,\text{ cm}^2$ at 1\,GeV/$c^2$ and masses as low as 73\,MeV/$c^2$~\cite{CRESST:2019jnq,cresst2024light,CRESST:2024cpr}. 
%Currently in its third phase, the experiment operates cryogenic calorimeters made from crystalline materials such as CaWO$_4$ \cite{CRESST:2020wtj}, Al$_2$O$_3$, Si \cite{PhysRevD.107.122003}, and LiAlO$_2$ \cite{CRESST:2022dtl} at approximately 15\,mK. Using highly sensitive transition-edge sensors (TESs) with detection thresholds as low as 6.7\,eV for gram-scale detectors \cite{CRESST:2024cpr} and 30.1\,eV for a 23.6\,g  crystal~\cite{CRESST:2019jnq}, CRESST has successfully probed sub-GeV DM with spin-independent DM-nucleon scattering cross sections down to $\mathcal{O}(10^{-37}\text{ cm}^2)$ \cite{CRESST:2019jnq,PhysRevD.107.122003,cresst2024light}. 
Further remarkable results that have been achieved both in terms of technological advancements and particle physics include single-photon detection capabilities using a silicon-on-sapphire (SOS) cryogenic detector~\cite{CRESST:2024cpr}, and the first measurement of the $^{180}$W $\alpha$-decay, with a half-life of $1.8(2) \times 10^{18}$ years \cite{cozzini2004detection}. 
In addition to the achievable thresholds, another notable advantage of the CRESST technology is its flexibility in the target material, allowing the use of nuclei sensitive to both SI and SD couplings \cite{CRESST:2022dtl,CRESST:2024cpr,CRESST:2019jnq}.

With well-established technology, CRESST is ready to embark on a comprehensive upgrade. 
This new phase will feature 288 readout channels, significantly increasing the achievable exposure, with the aim of unprecedented sensitivity in sub-GeV DM searches reaching a scattering cross section of $\mathcal{O}(10^{-42})~\mathrm{cm}^2$ at 1\,GeV/$c^2$. 
Although the primary goal of CRESST is to probe new DM parameter space for elastic DM-nucleus scattering in the sub-GeV range, it will also enable investigations into other intriguing physics cases, including searches for solar axions \cite{Abdelhameed:2020hys}, axion-like particles (ALPs) \cite{Mitridate2021}, dark photons \cite{CRESST:2016qpj} and constraints on self-interaction cross-sections of dark matter in universal bound states~\cite{CRESST:2024ahy}.
However, this upgrade presents several challenges. 
The large number of low-temperature detectors with TESs and DC-SQUIDs readout introduces additional complexity in the setup. 
A reformulation of the data processing pipeline is underway to handle the increased data rate and to automate the optimization of detector operational conditions including bias current using reinforcement learning \cite{Angloher:2023oya}.
First-level quality and live-time cuts, with substantial machine learning efforts are foreseen \cite{CRESST:machine_learning} along with the adaptation of existing simulation techniques to efficiently model the entire setup. 
The new approach will use the likelihood normalization method from the current background model while minimizing the time required for simulations, analysis, and model development~\cite{CRESST:2023thg}.

The original plans for this upgrade have been delayed by the discovery of an increased background event rate below approximately 200\,eV, which was first reported by CRESST-III in 2019. This phenomenon is now known as the Low-Energy Excess (LEE)~\cite{CRESST:2019jnq}.
%In 2019, CRESST-III was the first to report increased background event rates below about 200\,eV, a phenomenon now known as the Low-Energy Excess (LEE)~\cite{CRESST:2019jnq}, which has delayed upgrade plans. 
A comparable LEE was soon also reported by other low-threshold experiments employing different detection techniques and/or target materials~\cite{10.21468/SciPostPhysProc.9.001,10.21468/SciPostPhysProc.12.013,Baxter:2025odk}. 
The scientific community is making considerable efforts to understand and mitigate this excess, which hampers sensitivity to light DM. 
To address the LEE, the collaboration has developed new detector layouts~\cite{GRAPES-3:2024yym,karl:2022, CRESST:DetDev2023}, launching dedicated measurement campaigns, the most recent of which began in 2024. 

In this paper, we provide a historical overview of the CRESST experiment over the past 25 years, a brief description of the cryogenic technology, and the latest results (Secs. \ref{sec:cresst_experiment} to \ref{current_status}). 
We then discuss the upcoming CRESST upgrade, including projections exploring various LEE mitigation strategies and threshold reductions (secs. \ref{sec:upgrade} and \ref{sec:postupgrade}). 
We conclude with the discussion of the discovery potential of DM and the prospects of reaching the solar neutrino background, along with a brief overview of post-upgrade plans (Sec.~\ref{sec:longterm}).

\section{The CRESST experiment}
\label{sec:cresst_experiment}

In 1993, the Technical University Munich and the Max-Planck Institute for Physics initiated a collaboration to develop a DM search experiment called ``Munich Dark Matter Search" using cryogenic calorimeters with TESs at LNGS~\cite{cresst-history}. 
In 1996, the experiment was renamed ``CRESST" as the collaboration expanded to include teams from the University of Oxford and LNGS.

The experiment, which was first located in hall B of LNGS, performed its first DM search in 2000 using four Al$_2$O$_3$ detectors with a mass of ${\sim}260$\,g each. 
The results from this initial phase, later known as CRESST-I, set the best limits on spin-dependent and independent interactions for DM particles with a mass around 1\,GeV/$c^2$ at that time~\cite{Angloher:2002in}.
However, high backgrounds prompted the collaboration to work on an improvement plan. 
To actively distinguish between background and potential DM signals a double readout scheme was developed, marking the transition to CRESST-II, based on the use of scintillating crystals as targets for the DM interaction, paired with much smaller cryogenic calorimeters (light detectors) for the detection of the scintillation light emitted by the main crystal. 
Several different scintillating crystals were tested and the prominent interest at that time for DM in the 100\,GeV/$c^2$ mass range motivated the selection of CaWO$_4$ scintillating crystals, given the content of the heavy tungsten nuclei, a spin-independent scattering cross sections that scales with the square of the atomic mass number~\cite{cirelli2024darkmatter}. 
After moving the setup to hall A, the first results obtained in 2004 with two ${\sim}300$\,g CaWO$_4$ detectors showing a strong suppression of
non-nuclear recoil backgrounds were published ~\cite{Angloher:2004tr}.

Subsequent upgrades of the electronics system enabled the operation of several detector modules (scintillating crystal and corresponding light detector) simultaneously, with first results published in 2007~\cite{Angloher:2008wer} and a more extensive measurement campaign concluding in 2011~\cite{Angloher:2011uu}. 
This campaign suffered from background events due to alpha decays on the surfaces facing the detectors, which was addressed by redesigning the detectors' holding structure. 
The results of the detector with the best performance with this improved holding in CRESST-II were published in 2014~\cite{Angloher2014}. 

With increased theoretical interest in light DM candidates, CRESST shifted its focus in order to explore a region of the parameter space for elastic, spin-independent DM-nucleon scattering corresponding to WIMP masses below a few GeV/$c^2$~\cite{CRESST:2015djg}. 
This phase of the experiment, called CRESST-III, further optimized the detector layout using smaller CaWO$_4$ crystals to lower the detection threshold, and reduced the background by enhancing the radiopurity of the crystals. 
%As   increasing the amount of detectors and read-out channels. In addition, change to new electronics and DAQ were also done. 
%    this part above is more about current upgrade 
The first results with the optimized detectors that achieved a threshold of 30.1\,eV were published in 2019. 
This unmatched threshold for massive calorimeters extended the detection sensitivity to DM masses as low as 160\,MeV/$c^2$~\cite{CRESST:2019jnq}. 
However, in this phase the experiment's sensitivity was limited by the appearance of LEE events near the detection threshold. % as mentioned in Sec.~\ref{intro}. 
In CRESST-III, also other target materials were introduced, such as Al$_2$O$_3$ and LiAlO$_2$ (better suited for the search for light DM and spin-dependent interactions), and new detector designs were developed to investigate the origin of the LEE.
More details will follow in Sec.~\ref{current_status}.

The CRESST Collaboration is working towards a new phase, which is expected to explore new DM parameter space. 
More details about the planned upgrade stage are described in Sec.\ref{sec:upgrade}.

\subsection{Working principle}
Typically, CRESST utilizes a scintillating crystal as a target material, measuring both a heat/phonon signal and a scintillation light/photon signal. 
When a particle interacts with the target, nearly all the deposited energy (${\sim}$~90\%) is transferred to the crystal as phonons, a percentage that remains largely independent of the type of interaction. 
Therefore, the phonon signal represents the energy deposited by the interacting particle, though it does not provide information about the particle's identity. 
The scintillation light signal, however, is highly dependent on the interacting particles. 
For beta/gamma interactions, leading to electron recoils, the emitted light in CaWO$_4$ accounts for no more than 7\% of the observed energy. 
The scintillation light emitted for nuclear recoils is significantly reduced due to quenching, making it roughly an order of magnitude lower~\cite{Strauss:2014zia}. 
Both the phonon and light signals are simultaneously measured by separate thermometers, and the ratio between the energy deposited in these two channels, known as the light yield (LY), is determined.

\begin{figure}[ht]
\centering
\includegraphics[width=0.48\textwidth]{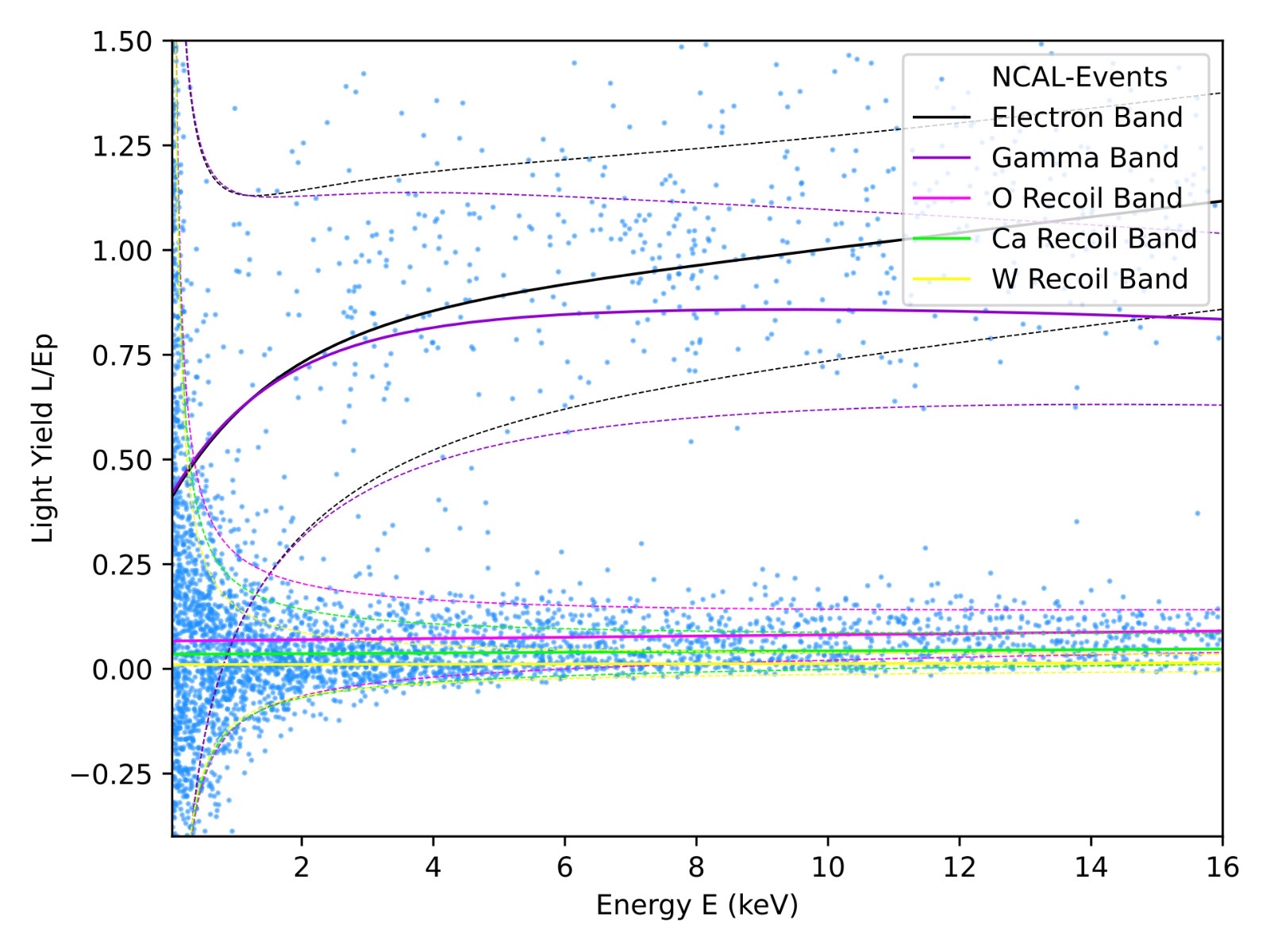}
\caption{The light yield (i.e. the energy of the light signal $L$ over the energy of the phonon signal $E_p$) versus energy plane for the neutron calibration data acquired with Detector A — a CRESST-III detector module with a CaWO$_4$ crystal target (taken from supplementary files of Ref.~\cite{CRESST:2024xwp}).}
\label{LY}
\end{figure}

The LY allows differentiation between particle interactions. 
As shown in Fig.~\ref{LY}, beta/gamma interactions exhibit the highest light yield, while the LY of nuclear recoils from Oxygen (O), Calcium (Ca), and Tungsten (W) is significantly lower. 
These recoils can be induced by either neutrons, neutrinos or DM particles. 
At energies above about 2\,keV, electromagnetic background events, primarily within the beta/gamma bands, are easily distinguishable from nuclear recoils. 
At lower energies, which are of particular interest for light DM searches, the beta/gamma bands (more information on the definition of bands can be found in ~\cite{CRESST:2024xwp}) overlap with the nuclear recoil bands, making it difficult to distinguish between different particle interactions on an event-by-event basis. 
For this reason, CRESST is extensively working on the development of its background model~\cite{CRESST:2019oqe, CRESST:2023thg} to continuously improve the identification of background components on a statistical basis, maintaining high sensitivity for light DM.

\subsection{CRESST setup}\label{setup}
\begin{figure}[ht]
\centering
\includegraphics[width=0.48\textwidth]{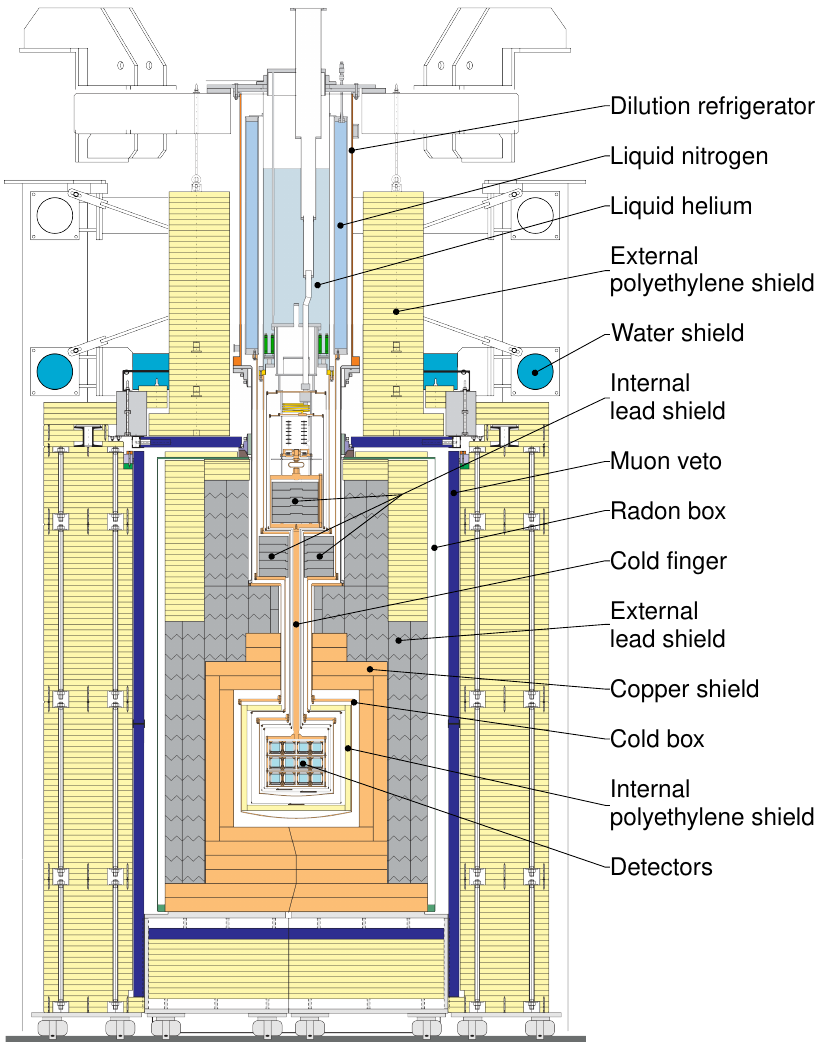}
\caption{\label{fig:setup} A schematic view of the CRESST setup.}
\end{figure}

The experiment is situated beneath the Gran Sasso massif. 
The surrounding rock provides a cosmic radiation shield equivalent to 3800\,meters of water~\cite{WULANDARI2004313}. 
To further reduce potential backgrounds, the experiment is protected by additional passive shielding and active veto layers. 
From the outermost to the innermost, these include a polyethylene (PE) layer, an active muon veto, a lead layer, a copper layer, and a second PE layer, as illustrated in Fig.~\ref{fig:setup}. The shielding layers inside the muon veto are enclosed in an air tight container (radon-box) which is constantly flushed with N$_2$ gas and maintained at a slight overpressure in order to prevent radon accumulation near the detectors. The PE shields against environmental neutrons. The muon veto, covering 98.7\% of the experiment's outer surface, identifies muons, allowing for the rejection of muon-induced events.
The experiment is shielded from gamma radiation by 20\,cm of lead, weighing 24\,t, which effectively absorbs gamma rays due to its high atomic number and density. 
Within the lead shielding, there is a 14\,cm thick layer consisting of 10\,t of radiopure copper layer. 
An additional PE layer inside the experimental volume shields the detectors from neutrons produced by interactions in the lead and copper. 
To minimize the radioactive background, all materials used around and within the detectors are carefully selected based on their radiopurity.

The cooling to the operational temperature of $\mathcal{O}(15)\,\mathrm{mK}$ is achieved using a commercial $^3$He/$^4$He dilution refrigerator. The cryostat and the dewars containing cryogenic liquids do not extend into the low background experimental volume (cold box), ensuring that there is no direct line of sight between the non-radiopure dilution refrigerator and the detectors. 
The low temperature of the dilution refrigerator is brought into the cold box via a 1.5\,m long cold finger. The cold box consists of five concentric radiation shields that surround the experimental volume and the cold finger. The cold finger and the shields are made of radiopure copper.

\subsection{Detector module}
A typical CRESST detector module consists of a (scintillating) target crystal, a light detector, and a reflective and scintillating housing. 
Following an energy deposition in the target crystal, a phonon signal is detected using a TES, a thin tungsten film deposited on a thin layer of amorphous SiO$_2$ which is placed directly on the crystal. 
The TES is operated at approximately 15\,mK, where tungsten is in the transition to its superconductive state. 
When phonons are absorbed by the TES, they raise its temperature, causing a measurable change in resistance.
The resulting change in current through an inductance coil alters the magnetic flux, which is measured by a SQUID amplifier.
To detect the light emitted by the scintillating target crystal, a 0.4\,mm thin SOS wafer equipped with a TES is used as a cryogenic light detector, also read out by a SQUID amplifier.

\begin{figure}[ht]
\centering
\includegraphics[width=0.48\textwidth]{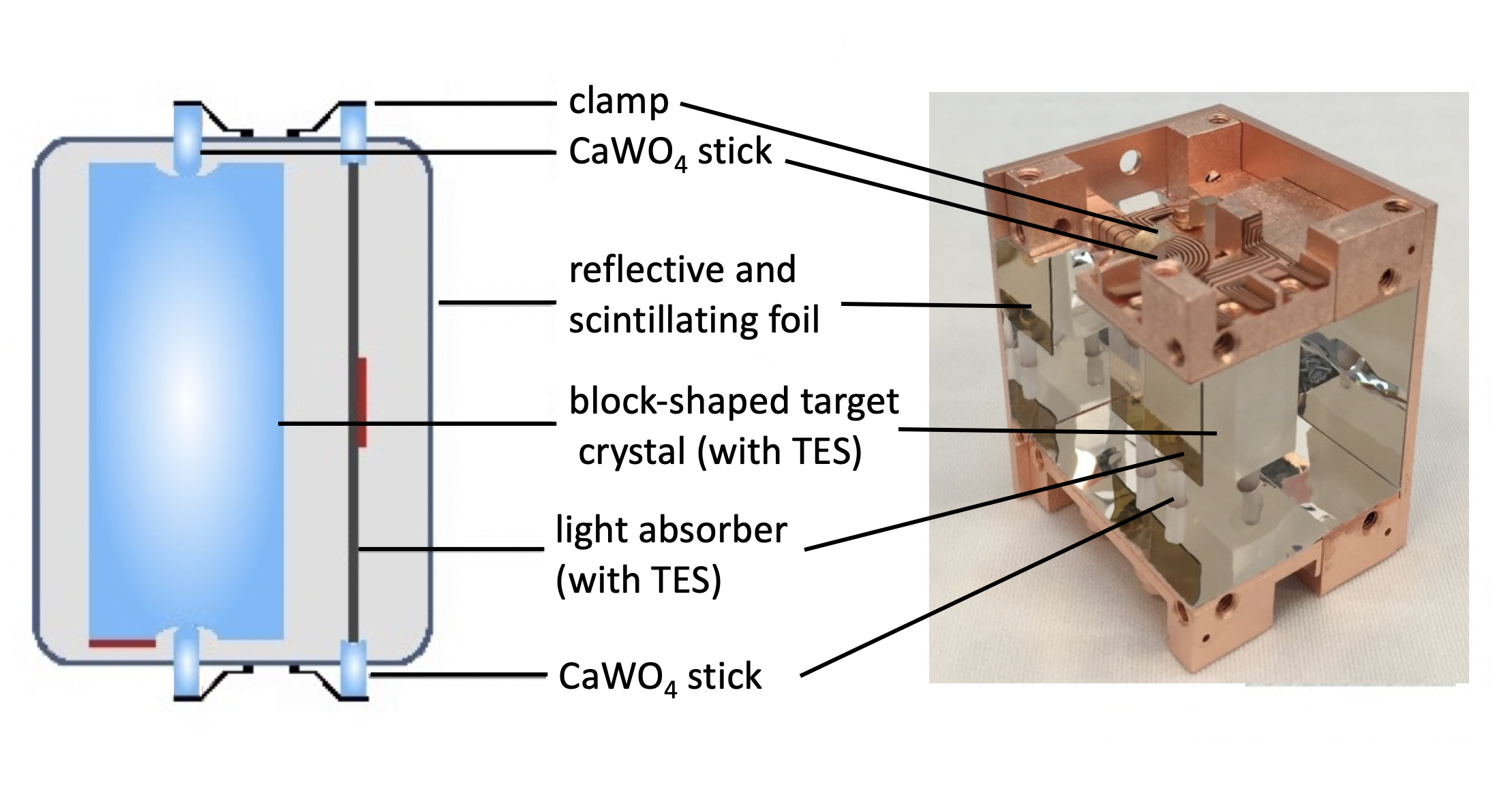}
\caption{Schematic view (on the left) and photo (on the right) of a CRESST-III detector module (not to scale). Parts in blue are CaWO$_4$, in red are the TESs. The block-shaped target crystal (20 x 20 x 10\,mm$^3$) has a mass of 23.6\,g. It is held by three CaWO$_4$ sticks. Three additional CaWO$_4$ sticks keep the light detector (20 x 20 x 0.4\,mm$^3$) in place.}
\label{fig:DetA}
\end{figure}

In the first run of CRESST-III (2016 - 2018), the detector modules were modified from those in CRESST-II, resulting in a detection threshold 10 times lower: Detector A, with a 23.6\,g block-shaped CaWO$_4$ crystal measuring 20\,x\,20\,x\,10\,mm$^3$, had by that time the lowest threshold of 30.1\,eV~\cite{CRESST:2019jnq}. 
%Both the target crystal and the light detector were held by three 12\,mm CaWO$_4$ sticks.
%The target crystal and the light detector were placed in a copper housing lined with reflective and scintillating Vikuiti$^{\circledR}$ foil, enhancing the light collection and creating, together with the CaWO$_4$ sticks a fully-scintillating surrounding for the detector. 
The target crystal and the light detector were held by three 12\,mm CaWO$_4$ sticks (each) and placed in a copper housing lined with reflecting and scintillating Vikuiti$^{\circledR}$ foil. This created a fully scintillating housing which allows to efficiently veto recoil events from alpha decays on surfaces or surface-near layers of materials surrounding the crystal~\cite{Angloher2014,CRESST:2014pmk}.
The detector and its schematic view are presented in Fig.~\ref{fig:DetA}.

Currently, the CRESST experiment tests in different runs a variety of detector modules, each exhibiting distinct configurations~\cite{10.21468/SciPostPhysProc.12.013,CRESST:DetDev2023}.  These variations include the use of target materials other than CaWO$_4$ in certain modules, as well as differing crystal holding mechanisms, applying either bronze clamps or copper sticks and gravity-based approaches. These modifications are systematically implemented to investigate potential sources of the LEE.

\section{Current status}\label{current_status}

\begin{figure}[ht]
    \centering
    \includegraphics[width=1.0\linewidth]{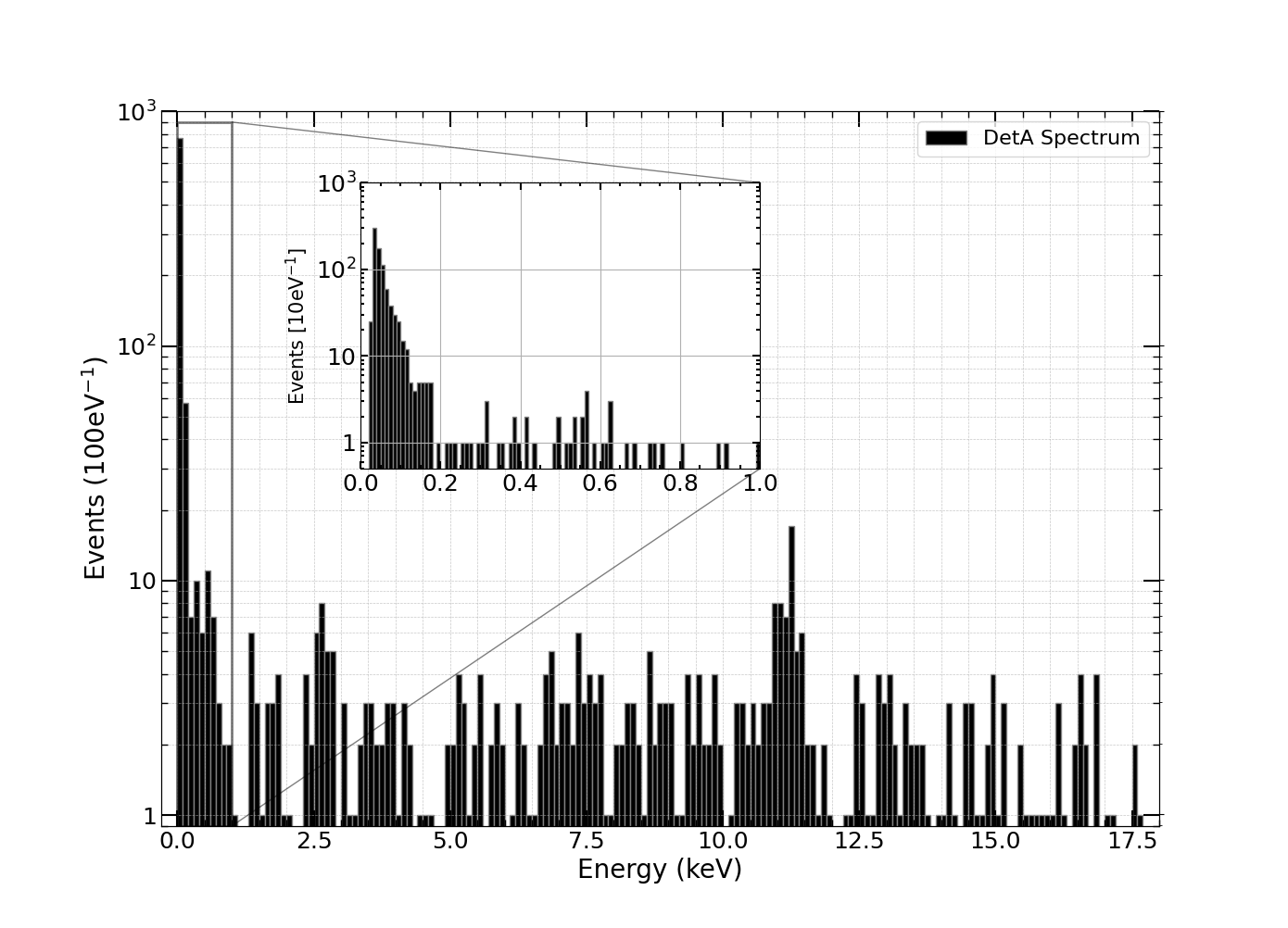}
    \caption{Energy spectrum of Detector~A~\cite{CRESST:2019jnq} with an inset plot showing the LEE. The lines at 2.6\,keV and 11.27\,keV are from cosmogenic activation of $^{182}$W~\cite{CRESST:2014irx}.} 
    \label{fig:LEEdetA}
\end{figure}

As mentioned in Sec.~\ref{intro}, the LEE refers to a steeply increasing event rate (below about 200\,eV in the CRESST data) towards decreasing recoil energies which has been observed in all cryogenic low-background experiments with a low enough threshold~\cite{10.21468/SciPostPhysProc.9.001,Baxter:2025odk}. 
This excess cannot be explained by a DM signal, known particle-induced background sources, or noise fluctuations large enough to trigger the detector \cite{10.21468/SciPostPhysProc.12.013,Baxter:2025odk}.
The CRESST experiment first observed the LEE during the CRESST-III operations in Detector~A \cite{CRESST:2019jnq}.
Other modules of the same run that achieved an energy threshold below 100\,eV also showed an LEE, but the focus in the following will remain on the detector with the lowest threshold, i.e. Detector~A.  
The energy spectrum observed in this detector is shown in Fig.~\ref{fig:LEEdetA}. 
The CRESST-III phase of the experiment achieved world-leading exclusion limits on the spin-independent DM-nucleon cross section below about 1\,GeV/$c^2$, reaching sensitivity to DM masses as low as 160\,MeV/$c^2$ \cite{CRESST:2019jnq}. 
More recently, CRESST set new world-leading limits below 0.2\,GeV/$c^2$, reaching down to a mass of 73\,MeV/$c^2$ \cite{CRESST:2024cpr}. 
However, in both cases, the LEE overlaps with a large part of the DM search region in the recoil energy spectrum, significantly reducing the sensitivity to the cross section as will be detailed in Sec.~\ref{sec:postupgrade}. %by up to two orders of magnitude. 
Therefore, the identification, characterization and mitigation of the LEE has become the highest priority in recent years.

\subsection{Investigating the low-energy excess}
The CRESST Collaboration has put significant effort into understanding the excess background through dedicated measurements. 
Different module designs were developed and tested, and focused analyses have been conducted, to isolate possible causes for this unexpected background, as will be laid out in this section. 
Tests have been carried out both at a surface facility and at the LNGS CRESST facility to avoid possible effects that are exclusive to measurements above ground or certain cryostats.  
The collective findings of these investigations show that the LEE is currently dominated by detector-intrinsic effects rather than a particle origin, aligning with the community-wide observations~\cite{10.21468/SciPostPhysProc.12.013}.

\paragraph{Detector layout studies: }
Extensive studies have been conducted with detector modules exploiting varying target materials, holding structures, and surrounding materials. 
Observations show that the LEE rate and spectral shape varies between detector modules and persists in non-scintillating materials, conflicting with an external origin such as luminescence, external radioactivity, or single-particle origins like DM.
The following specialized detector designs have provided these crucial insights into the nature of the LEE and/or will help to further narrow down the detector-intrinsic phenomena causing the majority of the excess and enable vetoing of LEE events while maintaining a high signal efficiency~\cite{CRESST:DetDev2023, GRAPES-3:2024yym, karl:2022}:
\begin{enumerate}
    \item \textbf{DoubleTES Detectors}: These modules employ two identical TESs, on the target crystal to distinguish between events occurring within the absorber bulk and those at or near the TESs, plus a light detector. Initial studies reveal that a component of the LEE originates at or near one of the TESs, pointing towards interface effects as a source.
    \item \textbf{Mini-Beaker Module}: This module features a $4\pi$ veto and an instrumented holding structure designed to tag surface radiation and events from the surroundings. The cylindrical $\text{Al}_2\text{O}_3$ target crystal is surrounded by a silicon beaker and an Al$_2$O$_3$ ring, all instrumented with TESs. The veto channels significantly improve background tagging.
    \item \textbf{Centimeter-Cube (cm-Cube) Array Module}: This design uses an array of four 1\,cm$^3$ $\text{CaWO}_4$ crystals with a low-force holding scheme and two light detectors, one above and one below the absorbers. The small absorber size reduces energy thresholds, enhancing sensitivity to low-energy events. This module also allows for to study coincidences between the events. %While this module has not been part
\end{enumerate}

\paragraph{Time-dependence studies: }
The LEE rate has been observed to decay over time, exhibiting two distinct time constants. 
The faster decay component partially resets after thermal cycles — specifically, after warming the detectors to tens of Kelvin — excluding a particle origin as the main source.
Instead, a viable hypothesis that we proposed~\cite{zema2022excess} and are testing via a DoubleTES module is that a mismatch between the thermal expansion coefficients of the crystal and the sensor induces shear stress at the interface, storing energy which can eventually be released contributing to the low energy excess events. Also other collaborations consider this a plausible hypothesis, see e.g.~\cite{Anthony-Petersen:2022ujw}.
In general, this time dependence suggests that the LEE rate can be significantly reduced by maintaining stable, long-term cryogenic conditions~\cite{10.21468/SciPostPhysProc.12.013}. 
Above-ground detector tests reveal decay times for the LEE consistent with those observed underground, indicating a common origin that is independent of the location~\cite{GRAPES-3:2024yym}.

\paragraph{Pulse shape studies: }
Pulse shape analyses confirm that LEE events exhibit pulse shape characteristics indistinguishable from particle recoil events at the level of accuracy achievable with current technology~\cite{10.21468/SciPostPhysProc.12.013}.
This observation, together with the fact that the excess extends well above the energy threshold, rules out triggered noise fluctuations as the cause~\cite{Stahlberg:2021gqi}.

\subsection{Mitigation strategies and ongoing research}
Building on the insights gained about the LEE thus far, mitigation strategies are being developed at the detector design, operations, and analysis levels.
These efforts include but are not limited to an optimized choice of materials at the detector interfaces, %slow (adiabatic) cooldowns,the 
discrimination between events occurring in the bulk of the detector and at the sensor interface, and the inclusion of the time dependent behavior of the LEE in the likelihood analysis.
We also envision new detector designs optimised to match the thermal expansion coefficients of the interfaced materials. 
In parallel, investigations continue to deepen the understanding of the underlying physical processes, aiming to eliminate their cause. If it is not possible to fully eliminate the LEE, it is needed to identify a physically motivated analytical description that can be incorporated into the likelihood.
%
%The mitigation of the LEE is a key focus for the CRESST Collaboration, with efforts spanning from dedicated data analyses to the development of new advanced detector modules to investigate the LEE's origins and characteristics. 
%The latter aims at distinguishing detector-intrinsic effects from external contributions while maintaining a high efficiency for a potential signal.
As a last resort, a particularly simple and practical backup strategy takes advantage of the natural decay of the LEE rate over time.
Projections suggest that approximately 450~days after the initial cool-down of the experimental setup, the LEE rate will have decreased by a factor of 10. 
A factor 100 reduction would be reached after 900~days.
Previous long-duration runs of the CRESST experiment, along with other collaborations using comparable infrastructures (e.g. ~\cite{nature_2022}), have established the feasibility of multi-year continuous cryostat operation at base temperature.
This ``waiting-time" strategy provides an effective means to reduce background levels even in the worst-case scenario where no further advancements in LEE mitigation are achieved, despite the tremendous progress in recent years.

\section{Upgrade plans}
\label{sec:upgrade}

The ongoing CRESST efforts to enhance the DM sensitivity are focused on three primary developments: (i) improving the detector performance by achieving lower energy thresholds, (ii) exploring strategies to mitigate or eliminate the LEE, and (iii) increasing the exposure. 

Currently, there are 24 working readout channels in CRESST, allowing the simultaneous operation of 12 standard CRESST modules or 8 DoubleTES modules. 
This strongly limits the achievable exposure. 
To exploit the increased sensitivity of future detector modules, larger exposures and thus an increased amount of readout channels are required.
%Currently, the exposure is constrained by the number of readout channels available in the CRESST setup. 
%The current configuration accommodates 24 of the 33 available channels , corresponding to 12 standard CRESST modules or 8 DoubleTES modules. 
%To enhance the sensitivity through increased exposure, a greater number of detector modules  — and thus readout channels — is required. 
Combined with advancements in TESs, SQUIDs, and cabling even greater improvements are anticipated.
To achieve a significant increase in sensitivity, the CRESST Collaboration foresees an upgraded readout system with 288\,channels.
The experimental facility, as described in Sec.~\ref{setup}, particularly its dilution refrigerator unit, is capable of supporting this upgrade. 
The upgraded configuration incorporates 288 DC-SQUIDs, installed on the 4.2\,K flange within the liquid helium bath of the CRESST cryostat.
%Cabling to the inner vacuum will be routed via specially designed feedthroughs. 
%The detailed implementation of the cabling and integration into the cryostat is beyond the scope of this paper.

The entire data acquisition system and associated electronics will be upgraded, while maintaining the mechanical components of the setup. 
The collaboration has been developing an integrated detector operation and data acquisition system specifically for TES-based cryogenic detectors. 
These upgrades will provide the CRESST facility with state-of-the-art equipment, combining modern technological advancements with its robust and well-characterized infrastructure.

The detector layouts described in Sec.~\ref{current_status}, developed to study the origin of the LEE and to identify strategies for its mitigation, will also form the basis for the layouts for the next generation of CRESST. 
%Two types of modules (DoubleTES and Mini-Beaker) are currently operated in the  CRESST measurement campaign, which started in 2024 \cite{CRESSTwebsite}. 
Due to the enhanced capability of the DoubleTES design to reject the component of the LEE  originating from events in a single TES sensor, it is currently foreseen as the future baseline design. 
Further design optimizations are under consideration to enhance the sensitivity to sub-GeV DM while minimizing the LEE.  

The combination of these efforts is expected to lead to an improved sensitivity to light DM. 
In Sec.~\ref{sec:postupgrade} the sensitivity projections based on Detector~A under the assumption of different LEE conditions, thresholds and exposures will be presented  and discussed.

\section{Performance studies with upgraded setup and with LEE}
\label{sec:postupgrade}
\begin{figure}[ht]
\centering
\includegraphics[width=0.48\textwidth]{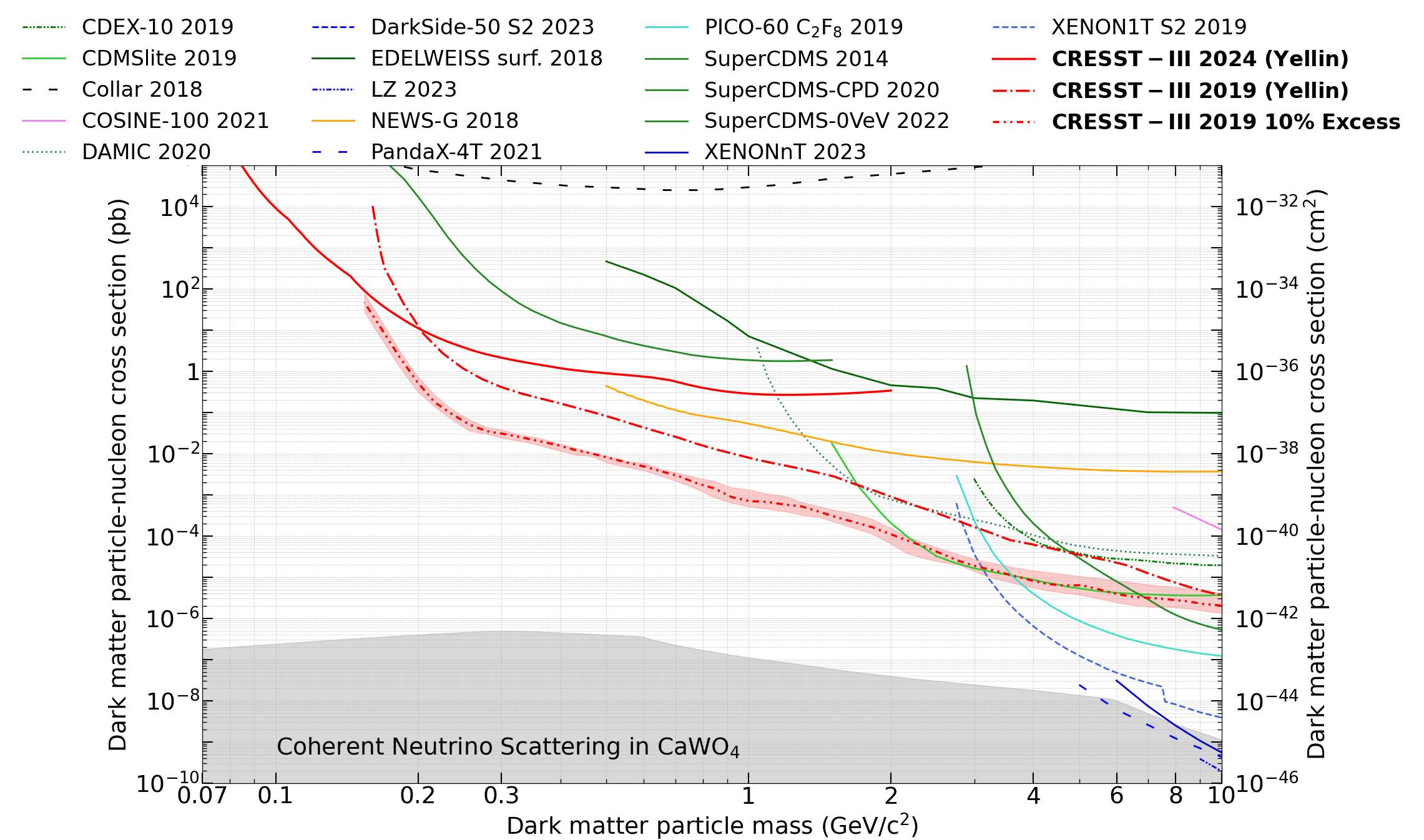}
\caption{\label{fig:LEEImportance} The experimental results for elastic, spin-independent DM-nucleus scattering are presented in the cross section versus DM particle mass plane. Unless explicitly stated otherwise, the results are reported with a 90\% confidence level (C.L.). This work presents CRESST-III results in red: with SOS as solid line~\cite{CRESST:2024cpr}, with CaWO$_4$ (Detector A) as dash-dotted line~\cite{CRESST:2019jnq}, with SOS as solid and the median of a projection of it with an assumption of only 10\% of the LEE %, which can be achieved even in the case of no LEE mitigation, 
as dash double dotted line. 
The band represent the $1\sigma$ region.
Color coding is used to categorize the experimental results: green represents exclusion limits (CDEX~\cite{CDEX:2019hzn}, CDMSlite~\cite{SuperCDMS:2018gro}, DAMIC~\cite{DAMIC:2020cut}, EDELWEISS~\cite{EDELWEISS:2016nzl}, SuperCDMS-CPD~\cite{SuperCDMS:2020aus}, SupersCDMS-0VeV~\cite{SuperCDMS:2022zmd}) obtained using silicon- or germanium-based solid-state detectors. Blue corresponds to liquid noble gas experiments using argon or xenon (DarkSide~\cite{DarkSide-50:2022qzh}, LZ~\cite{LZ:2022lsv}, Panda-X~\cite{PandaX-4T:2021bab}, XENONnT~\cite{XENON:2023cxc}, XENON1T~\cite{XENON:2019gfn}). Black dashed corresponds to organic scintillators - J. I. Collar~\cite{Collar:2018ydf}, orange solid denotes the gaseous spherical proportional counter NEWS-G (Ne + CH$_4$)~\cite{NEWS-G:2017pxg}, cyan represents the superheated bubble chamber experiment PICO (C$_3$F$_8$)~\cite{PICO:2019vsc} and pink for NaI crystals - COSINE~\cite{COSINE-100:2021xqn}. The gray area highlights the so-called neutrino fog, calculated for CaWO$_4$ in~\cite{bento2024solar}.}
\end{figure}

The presence of the LEE has significantly limited the experiment’s sensitivity to light DM, which is the primary focus of CRESST-III.
To mitigate this excess and to increase the potential discovery to light DM, the collaboration is taking the two aforementioned approaches: The LEE background is being reduced by implementing improved module designs, and advanced analysis strategies are being developed to mitigate the remaining background.
Based on recent advancements in detector development, an LEE reduction of a factor of 10 to 100 is deemed achievable. 
If none of the mitigation efforts fully eliminate the LEE, or at least significantly reduce it, the reduction factors of 10 and 100 are projected to be reached after approximately 1.5 and 3 years of continuous operation at base temperature, respectively.
These two reduction factors are considered as benchmark values for CRESST upgrade sensitivity studies. 
%A limit projection with the reduction factor of 10 for the LEE and assuming the performance of Detector A~\cite{CRESST:2019jnq} and of a SOS detector~\cite{PhysRevD.107.122003} with energy thresholds of 30.1\,eV and 6.7\,eV, respectively, was calculated using a likelihood framework~\cite{CRESST:2024xwp,Einfalt:2024rbt}.
A limit projection with the reduction factor of 10 for the LEE, while keeping the background level of Detector A~\cite{CRESST:2019jnq} constant, was calculated using a likelihood framework~\cite{CRESST:2024xwp,Einfalt:2024rbt}. The projection inherits the same 5.594\,kg$\cdot$day exposure as Detector A. In order to keep the spectral shape of the simulated data close to the observed spectrum, a simultaneous fit of the background and LEE is performed based on the method described in~\cite{CRESST:2024xwp,Einfalt:2024rbt}. The LEE is modeled with an exponential function. The reduction is simulated by scaling the amplitude of the LEE, while all other backgrounds remain unchanged. A thousand  Monte-Carlo simulations were performed from which the median and $1\sigma$ band were extracted.
%In this calculation, the performance of Detector A~\cite{CRESST:2019jnq} and of an SOS detector~\cite{CRESST:2024cpr} with energy thresholds of 30.1\,eV and 6.7\,eV, respectively, was assumed and the exposure was kept at 3.64\,kg·days.
The result is shown in Fig.~\ref{fig:LEEImportance} in comparison  with the current experimental status of world-wide elastic, spin-independent DM-nucleon scattering searches. 
The implementation of a likelihood analysis framework \cite{CRESST:2024xwp} has already demonstrated enhanced sensitivity even in the presence of the LEE and without including its time dependence.
Incorporating a time-dependent LEE model in the likelihood has the potential to further tighten the limits %, enhancing the discovery potential
for sub-GeV DM even in the absence of any other excess mitigation.
%The projections matt were calculated using a likelihood framework developed by CRESST~\cite{Einfalt:2024rbt} and are based on the performance of Detector A~\cite{CRESST:2019jnq} and of a SOS detector~\cite{PhysRevD.107.122003} with energy thresholds of 30.1\,eV and 6.7\,eV, respectively. 

%Assuming the CRESST modules will retain the double TES configuration, the channel availability after the setup upgrade will allow the operation of a total of 96 DoubleTES detector modules.  
%The projections are derived for an array comprising 70 modules equipped with $\sim$24\,g CaWO$_4$ crystals with a threshold of $\sim$30\,eV and 26 lighter modules of $\sim$2\,g crystals with a threshold of $\sim$5\,eV which are realistic and achievable performance targets for these modules, given the benchmark values obtained with similar detectors already operated in CRESST. The background of the CaWO$_4$ material assumed in the projections is the same as measured for the material of Detector~A.
%Acquiring data throughout the period of 3 years will result in an expected exposure of up to 1.5\,tonne·day.
%However, assuming Detector~A background levels, the experiment will be background limited after about 1\,tonne·day.
%For this reason, an exposure of 500\,kg·day is conservatively applied for the presented limit projections, obtained after about one year of data taking.

%As shown in Fig.~\ref{fig:proj}, the current exclusion limit can be significantly improved with a reduction of the LEE, particularly for light DM models. 
%Exploiting the fully improved setup will provide a substantial enhancement in sensitivity to DM-nucleon cross sections across all accessible DM masses.

\subsection{Primary science potential of the upgraded setup}
\label{sec:FutureLEE}
In the following subsections, potential configurations and design choices are being discussed for the detectors in the upgraded CRESST setup. This includes a consideration of various absorber materials, module arrangements, and readout channel allocations, all aimed at optimizing the setup’s performance for enhancing the overall sensitivity to DM particles.

\subsubsection{Spin-independent dark matter interaction}

As described above, the upgraded setup will include 288 readout channels. 
Assuming the CRESST modules will retain the DoubleTES configuration, each module requires two readout channels for the target crystal and one for light detection, accommodating up to 96 DoubleTES detector modules. 
Limit projections for spin-independent DM interactions are derived for a representative array comprising 70 modules equipped with 24\,g CaWO$_4$ crystals with a threshold of 30.1\,eV and 26 lighter modules of 2\,g CaWO$_4$ with a threshold of 5\,eV which are realistic and achievable performance targets for these modules, given the benchmark values obtained with similar detectors already operated in CRESST.
The background of the CaWO$_4$ material assumed in the projections is conservatively assumed to be the same as measured for the material of Detector~A.
Acquiring data throughout the period of 3 years will result in an expected total exposure of up to 1.5\,tonne·day.
However, assuming Detector~A background levels, the experiment will be background limited after about 1\,tonne·day independent of the LEE.
For this reason, an exposure of 500\,kg$\cdot$day is conservatively applied for the presented limit projections, obtained after about one year of data taking with the 70\,larger modules and a live time of 80\%.
The 26 lighter modules are designed to achieve lower energy thresholds and to enhance the sensitivity to light DM. 
When operated alongside the 70 larger modules, these smaller detectors will contribute an additional annual exposure of about 15\,kg$\cdot$day.

\begin{figure}[ht]
\centering
\includegraphics[width=0.48\textwidth]{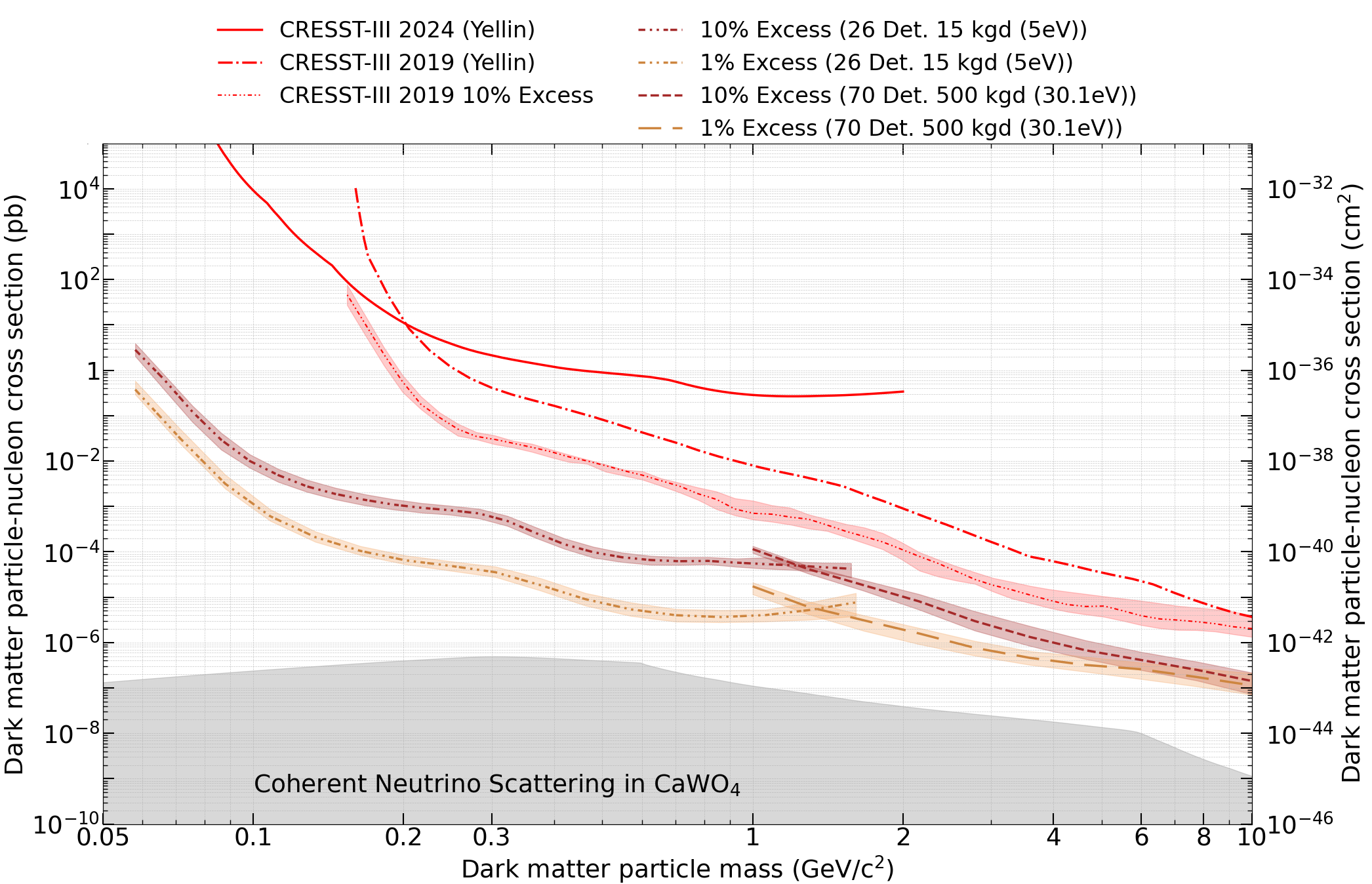}
\caption{\label{fig:proj} Projected sensitivities at 90\% C.L. on the elastic, spin-independent DM-nucleon scattering cross section based on the Detector~A performance exploring 26 CaWO$_4$ detectors with a threshold of 5\,eV and an exposure of 15\,kg$\cdot$day (double dash double dotted) and 70 CaWO$_4$ detectors with a threshold of 30.1\,eV and an exposure of 500\,kg$\cdot$day (dashed) for two reduction scenarios of the existing Detector A LEE (Excess):  10\% (in light burgundy color) and 1\% (in light orange color). For comparison, the limit for Detector~A~\cite{CRESST:2019jnq} (dash dotted red line), the same limit but with 10\% LEE (dash double dotted red line), and the limit for the SOS detector~\cite{CRESST:2024cpr} (solid red line) are shown. The bands represent the $1\sigma$ region around the median. The gray area highlights the so-called neutrino fog, calculated for CaWO$_4$ in~\cite{bento2024solar}.}
\end{figure}

Fig.~\ref{fig:proj} presents the resulting limit projections for both sets of modules and for LEE reduction factors of 10 and 100, as motivated earlier. % based on a combined exposure of 515\,kg·day. 
%These projections rely on the performance of the benchmark detectors already operated in CRESST in terms of background levels, energy resolution, and threshold. 
%The LEE is assumed to be mitigated by factors of 10 and 100, as discussed above. %reductions deemed achievable given recent advancements in detector development. 
%In the worst case scenario, achieving a factor ten reduction in the LEE rate would require a waiting period of approximately 450\,days, while a factor 100 reduction would require 900\,days, in the absence of any advancement in LEE mitigation. Previous long-duration runs of the CRESST experiment, along with other collaborations using comparable infrastructures (e.g. Ref.~\cite{nature_2022}), have established the feasibility of multi-year continuous cryostat operation at base temperature. Including a time-dependent LEE model in a likelihood analysis is expected to further enhance the DM sensitivity. Here, the 
The LEE background was modeled using a single energy-dependent exponential function constant in time. 
The energy-dependent signal survival probability in Detector~A was determined during data analysis. 
To obtain a realistic estimate for the 2\,g modules, the signal survival probability of Detector A was scaled to account for the lower threshold. 
For simplicity, identical cut efficiencies, acceptance regions, and energy resolutions were assumed across all detectors. 
The energy resolution at zero energy is set to one fifth of the threshold value for each module.  
The number of simulated events is scaled according to the respective exposure and follows a Poisson distribution. 
The strength of the LEE is parameterized by a multiplicative factor applied to the exponential function describing the LEE. %Exclusion limits are determined using a profile likelihood approach. 
%In the final exclusion limit, for each DM mass, only the best-performing detector type is considered. 
For masses up to about 1\,GeV/$c^2$, the low-threshold 2\,g modules with an exposure of 15\,kg$\cdot$day provide the strongest exclusion sensitivity.
These specialized detectors extend the experiment’s sensitivity range to cover DM masses below 60\,MeV/$c^2$, significantly broadening the experimental reach. 
At higher masses, the large combined exposure of the 24\,g modules yields superior exclusion power.
%We do not show the projections for the higher threshold detectors below masses of x GeV/c2 to keep the plot more readable
To ensure the readability of the plot, the projections for the two sets of modules are shown only in the mass range where they exhibit leading sensitivity.

\subsubsection{Spin-dependent dark matter interaction}
The CRESST experiment has previously demonstrated the sensitivity of lithium-based targets to spin-dependent interactions. 
Lithium, with its two isotopes, $^6$Li and $^7$Li, is particularly well-suited for these studies. % due to its one to two unpaired nucleons, 
It features unpaired protons and unpaired neutrons, making it sensitive to spin-dependent interactions with both protons and neutrons, and the model uncertainties on the nuclear structure with only few nucleons are comparatively small.
In an earlier run, CRESST successfully operated two LiAlO$_2$ detector modules \cite{CRESST:2022dtl}.
This configuration achieved leading limits on spin-dependent proton-only interactions for DM particle masses in the range of 0.25 to 2.5\,GeV/$c^2$ showcasing the potential of lithium-based targets in the search for light DM.  In the range of 0.074 to 0.25\,GeV/$c^2$, CRESST SOS detectors~\cite{CRESST:2024cpr} obtained leading limits, using $^{27}$Al as a probe to neutron-only  interactions.

Building on this success, the upgraded CRESST setup will allow for the operation of a significantly larger number of LiAlO$_2$ detectors, substantially improving the sensitivity to spin-dependent interactions.  
The main background source for these detectors is the decay of tritium, which is produced through neutron capture on $^6$Li. 
Reducing this background will require significant efforts to prevent the lithium from exposure to neutrons from mining extraction to crystal production and detector operation. 
This background, however, is well understood \cite{CRESST:2022dtl} and can be incorporated into a likelihood analysis. 
Consequently, similar improvements as in the case of searches for spin-independent interactions in CaWO$_4$ are expected.

%Figure~\ref{fig:SDLEE} illustrates the sensitivity projections for spin-dependent interactions with neutrons and protons, assuming the use of LiAlO$_2$ detectors. The projections are based on a total exposure of 3.8\,kg·days, corresponding to the operation of 13 detector modules, each with a mass of 10.46\,g.
%4\,cm$^3$
%These results highlight the potential of the upgraded CRESST setup to push the boundaries of current limits on spin-dependent DM interactions, particularly in the low-mass regime.  

%\begin{figure}[H]
%\centering
%\includegraphics[width=0.48\textwidth]{figures/SD_Limits.png}
%\caption{\label{fig:SDLEE} \textcolor{red}{will be updated if we will manage to run LIMITLESS} Exclusion limits set by various direct detection experiments for %spin-dependent interactions of DM particles in comparison with CRESST-Li (result~\cite{CRESST:2022dtl} and projections depicting 1, 10 and 100\% of LEE (Excess) %scenarios) with neutrons (upper plot, (CDMSlite~\cite{SuperCDMS:2017nns}, LUX~\cite{LUX:2016sci} and EDELWEISS~\cite{EDELWEISS:2019vjv}) ) and protons (lower %plot, (PICO~\cite{PICO:2017tgi}, LUX~\cite{LUX:2016sci}, J. I. Collar~\cite{Collar:2018ydf}, CDMSlite~\cite{SuperCDMS:2017nns} and %Borexino~\cite{Bringmann:2018cvk})).}
%\end{figure}

\subsection{Additional science potential with CRESST detectors}
\label{sec:alternative}

\subsubsection{Bosonic dark matter: dark photons and ALPs}

Relic dark photons and ALPs, collectively referred to as dark bosons, are expected to be absorbed in cryogenic calorimeters followed by the emission of an electron carrying the incoming energy~\cite{An:2014twa,CRESST:2016qpj, Hochberg:2016sqx}.
The expected signature of dark bosons is thus a mono-energetic electron recoil peak centered on the rest mass of the dark boson, with a width determined by the resolution of the detector. 
This spectral shape is distinct from the monotonically decaying energy spectrum of a potential LEE background.
CRESST-III achieved energy thresholds down to 6.7\,eV with a SOS detector~\cite{CRESST:2024cpr} and down to 10\,eV with a silicon detector~\cite{PhysRevD.107.122003} which correspond directly to the dark boson mass sensitivity. 
Preliminary studies based on the methods developed in~\cite{Zema:2024epe} demonstrate the feasibility of these searches in CRESST. 

%Initial studies ~\cite{Zema:2024epe} performed using ideal detectors (100\% detection efficiency), sapphire absorbers, and an exposure of 1\,kg·year, with a region of interest (ROI) of [9–100]\,eV, considering 90\% CL sensitivity limits obtained using  DarkELF~\cite{Knapen:2021bwg}, interfaced with the code developed shows the possible viability of these searches on CRESST.

%In the presence of the LEE an energy deposition from dark boson absorption would manifest itself as a distinct peak above a monotonically decaying energy spectrum. 
The future CRESST experiment has the potential to probe a new region of the bosonic DM mass range, specifically for dark photons and ALPs. 
To have a statistically significant observation of a peak above background, the larger exposure accessible with the upgraded setup will be required. 
%Initial studies were performed using ideal detectors (100\% detection efficiency), sapphire absorbers, and an exposure of 1\,kg·year, with a region of interest (ROI) of [9–100]\,eV, considering 90\% CL sensitivity limits obtained using  DarkELF~\cite{Knapen:2021bwg}, interfaced with the code developed and in~\cite{Zema:2024epe}.

\subsubsection{Solar axions}

The potential of cryogenic bolometers for axion detection is demonstrated in \cite{Abdelhameed:2020hys}, which presents an experiment utilizing a Tm$_3$Al$_5$O$_{12}$ crystal to search for solar axions.
This crystal contains $^{169}$Tm nuclei, which serve as targets for axion detection via resonant absorption.
Given the demonstrated sensitivity of this method and its scalability, it is worth exploring the feasibility of incorporating solar axion searches as a new research avenue within the CRESST program. 
The upcoming upgrade of the CRESST experiment is expected to increase the number of simultaneously operating detectors, offering a promising platform for such investigations. 

The Tm$_3$Al$_5$O$_{12}$ crystals would allow to simultaneously search for light DM and solar axions interactions opening the possibility of different DM-model studies inside CRESST. 
By utilizing the CRESST facility at LNGS, the extensive shielding that protects against environmental and cosmic radiation is likely to reduce background noise, thereby enhancing the sensitivity to solar axions. 
%Furthermore, $^{169}$Tm is an especially favorable target for axion detection, since unlike the $^{57}$Fe and $^{83}$Kr nuclei, the probability ratio $\omega_A / \omega_\gamma$ \cite{Derbin2023}.

\section{Long term vision}
\label{sec:longterm}

A long-term vision involves the possibility of using a large array of low-threshold cryogenic detectors in a next-generation rare event search. Such an experiment would have a rich scientific program and address several key scientific objectives, including direct DM detection, axion searches, and precision measurements of coherent elastic neutrino-nucleus scattering (CE$\upnu$NS). The key enabler for such a broad physics program is the development of low-threshold detectors (sensitive to energies of $\mathcal{O}(\text{1})\,$eV), the use of diverse nuclides as targets, high exposure, exceptional detector stability, and low background. These objectives require the low-threshold rare event search community to join efforts and they align closely with the overall goals of the future CRESST program.

\begin{figure}[ht]
\centering
\includegraphics[width=0.48\textwidth]{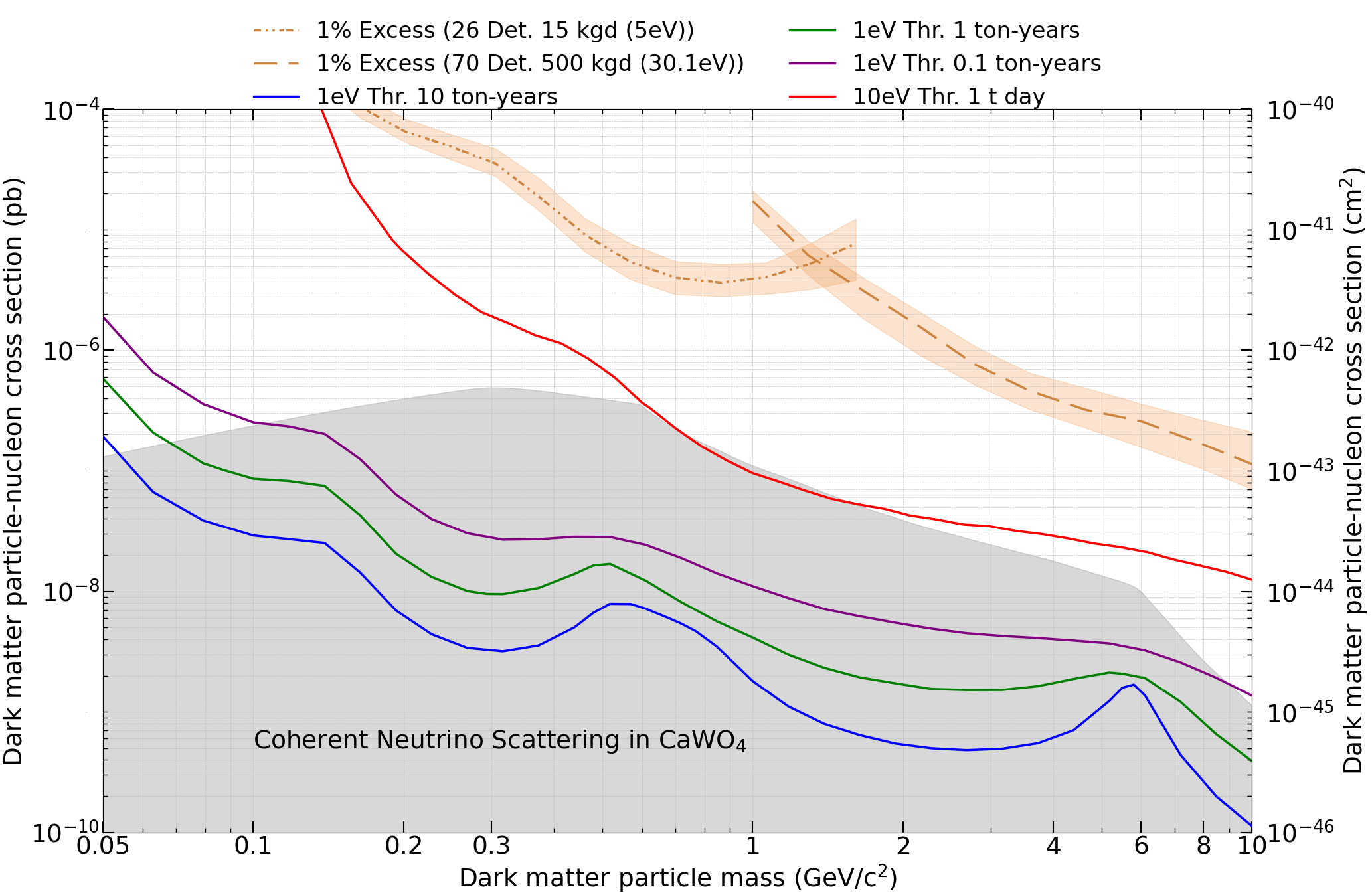}
\caption{\label{solar_neutrino} The four lower curves represent the $3\sigma$ discovery potential for spin-independent DM-nucleus scattering in the presence of solar neutrinos and no other backgrounds, assuming various exposures and thresholds. Below these curves, due to neutrinos presence a DM discovery becomes impossible. CRESST after the upgrade is expected to remain approximately two orders of magnitude above the neutrino fog, assuming detector thresholds of 30.1\,eV. %Achieving thresholds of 10\,eV with 23.6\,g detectors or 1\,eV with 2\,g detectors (currently at 6\,eV) is one of the primary goals for the next decade, while increasing exposure. 
The upper curve (orange) is the sensitivity projection for the CRESST upgrade assuming a 1\% LEE (Excess) with 515 kg$\cdot$day exposure (26 detectors of 2\,g with 5\,eV threshold (double dash double dotted) and 70 of 23.6\,g with 30.1\,eV threshold (dashed) in 1 year of data taking). The gray region highlights the so-called neutrino fog, calculated for CaWO$_4$ in~\cite{bento2024solar}.}
\end{figure}

Simulations conducted for the next-generation of cryogenic solid-state experiments \cite{bento2024solar} indicate that a 1\,tonne$\cdot$day exposure of CaWO$_4$ with no LEE and $\mathcal{O}$(10)\,eV threshold could start detecting solar neutrino events in the case of zero background as can be seen in Fig.~\ref{solar_neutrino}. 
From this figure, it is clear that CRESST will remain approximately two orders of magnitude above the neutrino fog if the background is not further reduced and the exposure increased.
Additionally lowering the threshold will further expand the accessible neutrino signal region. 
%a threshold of $\sim$30\,eV. 
%Reaching this region will require a further reduction in threshold by a factor of three and background reduction. %Although this appears to be a significant gap, in terms of threshold, it is reasonable to expect such progress within the next 5–10\,years. 
%In 2016, CRESST-II achieved a threshold of 600\,eV with 300\,g detectors \cite{Angloher2016}. By 2019, a threshold of 30.1\,eV was reached with the 23.6\,g Detector A~\cite{CRESST:2019jnq}. In 2024 CRESST-III has achieved thresholds as low as 6.7\,eV using light detectors \cite{CRESST:2024cpr}, emphasizing the continued progress in threshold reduction.
%However, even in the case of large background reduction, it will not be possible to reach the neutrino floor without increasing the exposure or reducing even more the threshold.  
Reaching a 1\,eV threshold together with a background reduction of the order of 10 within the next decade with this technology is a key objective of the CRESST Collaboration. 
In this scenario, neutrino events will start to play an important role in the background of CRESST.

The upcoming phase of  CRESST experiment is expected to reach a total exposure of 1\,tonne$\cdot$day after a three-year data-taking period. 
If DM is not discovered and CRESST proceeds with a follow-up upgrade with increased exposure, decreased background and lower thresholds, the experiment could probe the neutrino fog. 
To achieve this goal, future advancements in detector multiplexing, background mitigation, and threshold reduction will drive CRESST’s progress towards its long-term goals of being sensitive to DM in a region where neutrino interactions play an important role.

\section{Conclusion}

The CRESST Collaboration presents an ambitious yet feasible upgrade plan to design, hallbuild, and operate the experiment in the coming years. This new phase of the CRESST program will feature approximately 100 detectors, each with a threshold between about 5\,eV and 30\,eV, and plans an exposure of at least 500\,kg$\cdot$day. 
This configuration is expected to enable sensitivity to cross sections as low as $\mathcal{O}(10^{-43})~\mathrm{cm}^2$ and to DM masses below 100\,MeV/$c^2$.
%to sub-GeV DM with cross sections as low as $\mathcal{O}(10^{-43}~\mathrm{cm}^2)$.

While great progress has been achieved in elucidating the origin and pursuing the reduction of the LEE, these efforts are still ongoing, and further notable improvements are anticipated. 
But even in case the LEE cannot be fully eliminated, the CRESST experiment can deliver leading results.
Classification of events occurring exclusively at the interface between the target and a sensor has been demonstrated already with new detector module designs.
This enables the rejection of excess events that are localized at such interfaces, in contrast to DM events which are expected to interact uniformly throughout the target bulk. 
By employing a time-dependent LEE model in a likelihood-based framework, the impact of the LEE on the experiment's sensitivity can be even further reduced. 
Moreover, as the LEE decays over time, after 1.5 years of operation a tenfold reduction in the LEE is anticipated, providing a cross section sensitivity that improves over CRESST-III results by more than two orders of magnitude after at most an additional year of data-taking.
After three years of operation, a reduction factor of 100 is expected, further enhancing the sensitivity significantly.
%As previously discussed, While efforts are ongoing to understand and mitigate the LEE, the CRESST experiment can provide leading results in the light DM region. 
%After one year of operation, a significant reduction in the LEE is anticipated, with an expected reduction factor of 100 after three years. 
%By employing a time-dependent LEE model in a likelihood-based framework, even greater reduction factors may be achievable. 
The CRESST upgrade phase will thus be highly sensitive to light DM with masses below 100\,MeV/$c^2$ and will additionally have sensitivity to spin-dependent DM interactions, solar axions, dark photons and ALPs.
%With an exposure of 1.5\,tonne$\cdot$day, the cross section sensitivity at DM masses of a few GeV/$c^2$ will be about one order of magnitude above the neutrino fog.
%After this upgrade campaign, CRESST-IV will acquire around 1.5\,tonne$\cdot$day of exposure and be sensitive to sub-GeV DM below a 100 MeV/$c^2$ mass laying one order of magnitude above the neutrino-fog, with possibilities to spin-dependent DM interactions, solar axions, dark photons and ALPs.
%
If DM is not discovered during this program, the CRESST Collaboration plans to pursue additional upgrades to further improve the sensitivity to light DM, facilitate solar axion and relic dark boson studies, and enable precision measurements of solar neutrinos.

%Bosonic DM? 

\acknowledgments
This work has been funded in part by the Deutsche Forschungs\-gemeinschaft (DFG, German Research Foundation) under Germany's Excellence Strategy – EXC 2094 – 390783311 and through the Sonderforschungsbereich (Collaborative Research Center) SFB1258 ‘Neutrinos and Dark Matter in Astro- and Particle Physics’, by the BMBF 05A20WO1 and 05A20VTA, by the Austrian Science Fund (FWF) \url{http://dx.doi.org/10.55776/PAT1239524} and by \url{http://dx.doi.org/10.55776/I5420}. JB and HK were funded through the FWF project P 34778-N ELOISE. The Bratislava group acknowledges a partial support provided by the Slovak Research and Development Agency (projects APVV-15-0576 and APVV-21-0377).

\textbf{Author contributions}
The manuscript was produced by P.V.G, V.M. and B.v.Kr. Figures 5 and 6 were done by F.D., figure 7 by D.F. All authors have read and agreed to the published version of the manuscript.
Authors are listed alphabetically by their last names.

\textbf{Data availability}
This manuscript has no associated data or the data will not be deposited.

\textbf{Competing interests}
The authors declare no competing interests.

\textbf{Additional information}
Correspondence and requests for materials should be addressed to
P.V.G, V.M. and B.v.Kr.

%%%%%%%%%%%%%%%%%%%%%%%%%%%%%%%%%%%%%%%%%%%%%%%%

\bibliography{refs.bib}

\end{document}